\documentclass[a4paper,12pt]{article}
\usepackage{epsfig,amssymb,amsmath}
\usepackage{graphicx}
\usepackage{amsfonts}

\usepackage{tikz}
\usepackage{tkz-euclide}
\tikzset{->/.style = {decoration={markings,
mark=at position 1 with {\arrow[scale=2]{latex'}}},
postaction={decorate}}}
\tikzset{<-/.style = {decoration={markings,
mark=at position 0 with {\arrowreversed[scale=2]{latex'}}},
postaction={decorate}}}

\tikzset{circ/.style = {fill, circle, inner sep = 0, minimum size = 3}}

\tikzset{eqpic/.style={baseline={([yshift=-.5ex]current bounding box.center)}}}

\definecolor{mblue}{rgb}{0.2, 0.3, 0.8}
\definecolor{morange}{rgb}{1, 0.5, 0}
\definecolor{mgreen}{rgb}{0.1, 0.4, 0.2}
\definecolor{mred}{rgb}{0.5, 0, 0}

\begin{document}

\newcommand{\be}{\begin{equation}}
\newcommand{\ee}{\end{equation}}
\newcommand{\bea}{\begin{eqnarray}}
\newcommand{\eea}{\end{eqnarray}}
\newcommand{\bean}{\begin{eqnarray*}}
\newcommand{\eean}{\end{eqnarray*}}
\font\upright=cmu10 scaled\magstep1
\font\sans=cmss12
\newcommand{\ssf}{\sans}
\newcommand{\stroke}{\vrule height8pt width0.4pt depth-0.1pt}
\newcommand{\Z}{\hbox{\upright\rlap{\ssf Z}\kern 2.7pt {\ssf Z}}}
\newcommand{\ZZ}{\Z\hskip -10pt \Z_2}
\newcommand{\C}{{\rlap{\upright\rlap{C}\kern 3.8pt\stroke}\phantom{C}}}
\newcommand{\R}{\hbox{\upright\rlap{I}\kern 1.7pt R}}
\newcommand{\HH}{\hbox{\upright\rlap{I}\kern 1.7pt H}}
\newcommand{\CP}{\hbox{\C{\upright\rlap{I}\kern 1.5pt P}}}
\newcommand{\identity}{{\upright\rlap{1}\kern 2.0pt 1}}
\newcommand{\half}{\frac{1}{2}}
\newcommand{\quart}{\frac{1}{4}}
\newcommand{\pr}{\partial}
\newcommand{\bm}{\boldmath}
\newcommand{\I}{{\cal I}} 
\newcommand{\M}{{\cal M}}
\newcommand{\N}{{\cal N}}
\newcommand{\e}{\varepsilon}

\thispagestyle{empty}
\vskip 3em
\begin{center}
{{\bf \huge Integration Theory for Kinks and Sphalerons in One Dimension}} 
\\[15mm]

{\bf \Large N.~S. Manton\footnote{email: N.S.Manton@damtp.cam.ac.uk;
    ORCID: 0000-0002-2938-156X}} \\[20pt]

\vskip 1em
{\it 
Department of Applied Mathematics and Theoretical Physics,\\
University of Cambridge, \\
Wilberforce Road, Cambridge CB3 0WA, U.K.}
\vspace{12mm}

\abstract
{The static kink, sphaleron and kink chain solutions for a single scalar field
$\phi$ in one spatial dimension are reconsidered. By integration of the
Euler--Lagrange equation, or through the Bogomolny argument, one finds
that each of these solutions obeys a first-order field equation, an
autonomous ODE that can always be formally integrated. We distinguish
the BPS case, where the required integral is along a contour in the
$\phi$-plane, from the semi-BPS case, where the integral is along a
contour in the Riemann surface double-covering the $\phi$-plane, and
is generally more complicated. 
}

\end{center}

\vskip 150pt
\leftline{Keywords: Kink, Sphaleron, Integration}
\vskip 1em

\vfill
\newpage
\setcounter{page}{1}
\renewcommand{\thefootnote}{\arabic{footnote}}


\section{Introduction} 
\vspace{4mm}

A scalar field theory in one spatial dimension, with a single scalar
field $\phi$ and appropriate vacuum structure, is well-known to have stable
kink solutions interpolating between adjacent vacua \cite{Raj,book,Shn}.
It may also have unstable, sphaleron (or bounce) solutions,
representing a saddle point between false and true vacua
\cite{CC,KM,Ave,MR}. Additionally, such a theory
generally has spatially periodic, kink chain solutions, composed of
alternating kinks and antikinks \cite{MS}. All these static solutions obey a
first-order ODE, obtained as the first integral of the theory's
second-order Euler--Lagrange equation. The different solutions arise by
varying the constant of integration, as well as the underlying
potential $U(\phi)$. Here, we will be
concerned almost entirely with solutions of this first-order ODE,
which we refer to as the field equation. We will review the solutions,
some examples of which are widely known and others less so, focussing
on their analytic and algebraic form. We will also find the energies of
kink and sphaleron solutions, and the spatial period and energy
per period of kink chains. All the calculations reduce to
integrals of differentials on the $\phi$-plane or its branched
double-cover $\Sigma$, many of which simplify to elementary integrals.

We start with the energy functional for time-independent fields,
\be
E = \half \int_{-\infty}^{\infty} \left\{ \left( \frac{d\phi}{dx}
  \right)^2 + U(\phi) \right\} \, dx \,,
\label{energy}
\ee
where $U$ is real for all real $\phi$. The second-order
Euler--Lagrange equation,
\be
\frac{d^2\phi}{dx^2} - \half \frac{dU}{d\phi} = 0 \,,
\label{EulerLag}
\ee
has the first integral
\be
\left(\frac{d\phi}{dx}\right)^2 = U(\phi) + C \,,
\label{firstint}
\ee 
where $C$ is an arbitrary constant. It is convenient to absorb
$C$ into the potential $U$ (i.e. set $C=0$), as we are interested in
particular solutions. Variations of $C$ are accommodated by adjusting
$U(\phi)$. Taking the square root of (\ref{firstint}), we obtain
\be
\frac{d\phi}{dx} = \sqrt{U(\phi)} \,,
\ee
the first-order autonomous ODE that we call the field equation. Its
formal solution is
\be
x = \int \frac{d\phi}{\sqrt{U(\phi)}} \,.
\label{soln}
\ee
It is assumed that $U$ (with the constant $C$ absorbed) is positive
for at least some range of $\phi$, so that $x$ is real.
The further constant of integration in (\ref{soln}) encodes a shift of $x$,
i.e. a spatial translation. We will not be careful about this, so all
solutions appearing below can be spatially translated.

We will assume that $U(\phi)$ is a holomorphic or somtimes meromorphic
function of $\phi$, real along the real $\phi$-axis.
Simplest is where $U$ is a polynomial in $\phi$, but we will later
consider $U$ a power of $\cos\phi$. $U$ can
have zeros, at real or complex values of $\phi$. At a zero of odd
multiplicity, $\sqrt{U}$ has a branch point; we will therefore need
to consider the Riemann surface $\Sigma$ that is the branched
double-cover of the $\phi$-plane. $\Sigma$ is defined by the algebraic
equation
\be
V^2 = U(\phi) \,,
\ee
and on $\Sigma$, the differential
\be
\frac{d\phi}{V(\phi)} = \frac{d\phi}{\sqrt{U(\phi)}} 
\label{diff}
\ee
is well-defined. The field equation's formal solution (\ref{soln})
becomes
\be
x = \int \frac{d\phi}{V(\phi)} \,,
\label{solnV}
\ee
where the integral is along a suitable open contour on $\Sigma$ that
keeps $x$ real.

There is still a choice of the overall sign for the square
root (equivalently, the choice of starting point for the contour on
$\Sigma$). A reversal of the sign reverses the sign of the derivative of
$\phi$ with respect to $x$, which has the effect of a spatial reflection,
and turns a kink into an antikink.
The sign needs to be chosen in a consistent and smooth
way along the contour of integration, especially if the contour
encircles a branch point. When we later reconsider the Bogomolny
argument \cite{Bog}, we will not need to explicitly allow for the sign
choice $\pm\sqrt{U}$ (as is usually done), because it is implicit
in the choice of contour.

We can gain considerable insight into solutions by considering the integral
(\ref{solnV}) locally near a zero of $U$. Let us assume the zero is at
$\phi = 0$. Suppose first that the zero is simple, and $U(\phi) =
\phi$ in leading approximation. Then
\be
x = \int \frac{d\phi}{\sqrt{\phi}} = 2\sqrt{\phi} \,,
\ee
so $\phi(x) = \frac{1}{4}x^2$. We see that although $x$ is a double-valued 
function of $\phi$, and potentially imaginary, the solution $\phi(x)$ itself 
is real and has a smooth minimum. A smooth, real maximum occurs if 
$U(\phi) = -\phi$. More generally, if $U$ has only simple zeros, then the
differential (\ref{diff}) has no singularities on $\Sigma$ (as
becomes clear after a change of variable $\psi^2 = \phi - a$ near
a simple zero of $U(\phi)$ at $a$). The differential is
therefore holomorphic on $\Sigma$ -- it is an Abelian differential of the
first kind -- and $x$ is its integral from some fixed initial point
to the variable point $\phi$. Exceptions are if $U$ is a quadratic or linear
polynomial; in these cases the differential has a pole at infinity.

Suppose next that $U$ has a double zero,
so locally $U(\phi) = \phi^2$. Then the Riemann surface $\Sigma$ is not
branched at $\phi = 0$, and the differential (\ref{diff}) has a simple
pole there. More precisely, $\Sigma$ has two disjoint pieces,
where $V = \sqrt{U} = \phi$ and $V = \sqrt{U} = -\phi$, and the poles of the
differential have opposite residues on these pieces. The Riemann
surface construction has blown-up the node $V=U=0$ of the curve
$V^2 = U(\phi)$ into two ``places''. The local solution (\ref{solnV})
is
\be
x = \pm \int \frac{d\phi}{\phi} = \pm\log\phi \,,
\ee
or equivalently $\phi(x) = e^{\pm x}$. This is the typical exponential
growth of a global solution for large negative $x$, or exponential
decay for large positive $x$, and is called a short-range tail.
Finally, suppose $U$ has a zero of even multiplicity
$2k$, with $k>1$, so locally $U(\phi) = \phi^{2k}$. Then the differential 
(\ref{diff}) has a pole of order $k$, and its integral is
\be
x = \pm \int \frac{d\phi}{\phi^k} = \mp\frac{1}{(k-1)\phi^{k-1}} \,,
\ee
so $\phi(x)$ is an inverse (fractional) power of $ \pm (k-1)x$. In the simplest
case $k=2$, $\phi(x)$ is proportional to $\pm\frac{1}{x}$. This is the
basic, long-range tail behaviour of a solution.

If all the zeros of $U$ are double, or of higher even multiplicity,
then $\Sigma$ is globally two copies of the $\phi$-plane, $V$ is a
polynomial, and (\ref{diff}) is generally an Abelian differential
of the third kind, i.e. a meromorphic differential including simple
pole parts, on each of these copies. In certain cases, (\ref{diff})
is an Abelian differential of the second kind, i.e. meromorphic,
with vanishing simple pole parts. 

Many well-known solutions of field theory have the behaviours emerging
from this local analysis. The kink solutions in the standard $\phi^4$-theory
and $\phi^6$-theory connect two double zeros of $U$ (true vacua),
and have exponentially decaying tails. A special $\phi^8$ kink occurs
in a theory where $U$ has one quartic zero ($k=2$) and two double zeros.
Here the kink has one long-range and one short-range tail \cite{GE}.
It is important to distinguish double zeros of $U$ from zeros of
higher, even multiplicity. Lohe \cite{Loh}, and more recently Khare et al.
\cite{KCS}, Bazeia et al. \cite{BMM}, and Gani and collaborators
\cite{RMGC,GMB,Bli} have drawn attention to several examples.

A sphaleron is a solution $\phi(x)$ that, as $x$ increases, runs from
a double zero $\phi_0$ of $U$ (a false vacuum), via a simple zero of
$U$, back to $\phi_0$ on the other sheet of $\Sigma$. $\phi(x)$ is a smooth
quadratic function of $x$ near the simple zero, and its spatial derivative
has a node (a zero) here. The spatial derivative is the translation
zero-mode of the sphaleron, its zero-frequency mode of small
oscillation, so by the Sturm oscillation theorem, there is inevitably
a single mode of negative squared frequency with no node. This is
the unstable mode of the sphaleron.

Finally, there are kink chain solutions connecting adjacent, simple
zeros of $U$, with $U$ positive between them. The integral (\ref{solnV}),
determining $x(\phi)$, is finite over a minimal integration contour
connecting these zeros, so the contour returns on the second sheet
of $\Sigma$, and then continues on the first sheet, etc. The result
is a spatially periodic solution $\phi(x)$, whose period is a real
period of the Abelian differential (\ref{diff}). Simple zeros of $U$
are generic, so, if we allow the constant of integration $C$ to vary,
then for almost all $C$, it is spatially periodic solutions that occur. 
Like sphalerons, these kink chain solutions are generally unstable.

It is important to understand the genus $g$ of the Riemann surface $\Sigma$,
as this significantly affects the global nature of the differential
(\ref{diff}), and hence the integral solution (\ref{solnV}) of the
field equation. The genus depends critically on the multiplicities
of the zeros of $U$. In most of the familiar
examples of kinks, $U$ has sufficiently many double
zeros, or zeros of higher even multiplicity, that the genus reduces
to zero and the differential (\ref{diff}) is meromorphic on the
$\phi$-plane. Using partial fractions, the differential can then be
integrated, giving an implicit solution $x = G(\phi)$. In simple cases, this
can be inverted to give an explicit solution $\phi(x)$, but
inversion is not possible in general.

Conversely, if $U$ is a polynomial of degree $2n$ (or degree
$2n-1$), with only simple zeros, then the Riemann surface $\Sigma$
defined by $V^2 = U(\phi)$, double-covering the $\phi$-plane, is of genus
$n-1$ and the solution of the field equation is an integral of an
Abelian differential of the first kind. (The case $n=1$ is special --
here the genus of $\Sigma$ is zero but the differential acquires
simple poles.) For all $n$, the solutions are spatially periodic kink chains.
If $U$ has a double zero, and further simple zeros, then a
sphaleron can occur. The genus is reduced to $n-2$ because of the
double zero. Also, the differential is meromorphic, with at least one
pair of simple poles. If $U$ has two double zeros, with the remaining
zeros simple, then a kink solution can connect the double zeros.
For $n > 2$, the genus is reduced to $n-3$ and the differential has
at least four simple poles.

Various other possibilities occur if $U$ has more double zeros,
or zeros of higher multiplicity. A general statement depends on the
observation that any polynomial $U(\phi)$ has an essentially unique
decomposition $U(\phi) = P^2(\phi)Q(\phi)$ where $P$ and $Q$ are
polynomials and $Q$ has only simple zeros. The curve $V^2 = U(\phi)$
then has the same genus as the curve $v^2 = Q(\phi)$.

\section{BPS and Semi-BPS Kinks}
\vspace{4mm}

In this section we first review the Bogomolny argument \cite{Bog} that leads
to a simple understanding of kink solutions and their energies. We then
clarify what we mean by a BPS kink, and introduce the new notion of a
semi-BPS kink. There is also the notion of a semi-BPS sphaleron (a notion
independently studied by Izquierdo et al. \cite{Izq}) and of
a semi-BPS kink chain. These are discussed later. 

We assume at first that $U(\phi)$ is a polynomial. For the usual Bogomolny
argument to work, $U$ needs to be positive semi-definite, with two or
more distinct zeros. These zeros of $U$ are quadratic, or of higher even order
(otherwise $U$ would take negative values). Physically, they are vacua
of the field theory, as they minimise the energy. Let $\phi_-$ and $\phi_+$
be adjacent zeros, with $\phi_+ > \phi_-$. Then we expect there to be
a stable kink solution interpolating between $\phi_-$ as $x \to -\infty$ and
$\phi_+$ as $x \to \infty$. The kink is the minimal-energy static solution
with these boundary conditions, and to find it we perform the
following Bogomolny rearrangement of the energy
(\ref{energy}) (known much earlier in this one-dimensional context),
\bea
E &=& \half \int_{-\infty}^{\infty} \left\{ \left( \frac{d\phi}{dx}
\right)^2 + U(\phi) \right\} \, dx \nonumber \\
&=& \half\int_{-\infty}^{\infty} \left\{ \left( \frac{d\phi}{dx} -
\sqrt{U(\phi)} \right)^2 + 2\sqrt{U(\phi)} \, \frac{d\phi}{dx} \right\} dx
\nonumber \\
&=& \half\int_{-\infty}^{\infty} \left( \frac{d\phi}{dx} -
  \sqrt{U(\phi)} \right)^2 dx
+ \int_{\phi_-}^{\phi_+} \sqrt{U(\phi)} \, d\phi \,.
\label{Bogomanip}
\eea
The final integral depends only on the form of $U$ and the endpoints
of the integral, and not on the $x$-dependence of $\phi$. It is
therefore ``topological''. The energy is minimised when
the remaining integral vanishes, i.e. when the first-order field
equation
\be
\frac{d\phi}{dx} = \sqrt{U(\phi)}
\ee
is satisfied. For the kink solution, the positive square
root of $U$ should be selected, so that $\frac{d\phi}{dx}$ is
positive. The negative square root gives the antikink, the spatial reflection
of the kink, with opposite boundary conditions.

The field equation obtained by this Bogomolny argument is the same as that
obtained as the first integral of the Euler--Lagrange equation, except
that the constant of integration $C$ is automatically zero. The
argument also gives the formula for the kink energy
\be
E = \int_{\phi_-}^{\phi_+} \sqrt{U(\phi)} \, d\phi
= \int_{\phi_-}^{\phi_+} V(\phi) \, d\phi\,.
\label{Bogoenergy}
\ee
Here it is important that $U(\phi_{\pm}) = 0$, otherwise the
energy would diverge.

The Bogomolny argument guarantees kink stability, since the kink is the 
minimal-energy field configuration connecting $\phi_-$ to
$\phi_+$. There is also a local argument. The spatial derivative of
the kink solution is everywhere positive, as $\sqrt{U(\phi)}$ is
positive between $\phi_-$ and $\phi_+$. But this spatial derivative is
the translation zero-mode of the kink, a zero-frequency eigenfunction
of the second-variation of the energy. As this mode has no node
(zero), the Sturm oscillation theorem implies that for the kink, there
are no modes of negative squared frequency, i.e. modes of instability.

We now clarify the distinction between a BPS kink and a semi-BPS
kink. We call the kink BPS if the polynomial $U(\phi)$ has a square root
$V(\phi)$ that is itself a polynomial. The integral of $V(\phi)$ is
another polynomial $W(\phi)$, and
\be
U(\phi) = \left(\frac{dW(\phi)}{d\phi}\right)^2 \,.
\ee
$W$ is known as the superpotential of $U$, and it occurs as a
fundamental ingredient in supersymmetric theories \cite{Fre}.
The Riemann surface $\Sigma$ separates into two
copies of the $\phi$-plane in the BPS case. The differential
(\ref{diff}) is meromorphic and can be integrated using partial fractions
\cite{Har}. For a BPS kink, the energy (\ref{Bogoenergy}) simplifies to
\be
E = \int_{\phi_-}^{\phi_+} V(\phi) \, d\phi = W(\phi_+) - W(\phi_-) \,.
\label{BogoenergyW}
\ee
Note that an additive constant of integration in $W$ has no effect,
either on the field equation, or on the kink energy.

A kink still exists if the polynomial $U$ does not have a polynomial square
root, but has double (or higher even order) zeros at $\phi_-$ and
$\phi_+$ and is positive between. Such a polynomial can have real simple zeros
outside this range, or complex conjugate pairs of simple zeros, so
$\Sigma$ is a connected double-cover of the $\phi$-plane, branched at
the simple zeros (and also at any zeros of higher odd order). We
call this type of kink semi-BPS. The Bogomolny argument still works,
and there is a kink connecting $\phi_-$ and $\phi_+$ whose energy is
given by the formula (\ref{BogoenergyW}). However, the field
equation is more difficult to solve, because the differential
(\ref{diff}) is only properly defined on $\Sigma$. Also, the function
$W(\phi)$ is more difficult to determine. 
The semi-BPS case becomes complicated if $U$ has more than two simple
zeros, because $\Sigma$ then has positive genus, and the
kink and its energy depend on elliptic or hyperelliptic
integrals. Note that if $U$ is negative for some range of real
$\phi$, then a semi-BPS kink is only metastable. 

\section{Examples of BPS Kinks}
\vspace{4mm}

The canonical example of a potential $U$ admitting a BPS kink is
the double-well potential of $\phi^4$-theory (Fig.1) \cite{DHN,Pol}
\be
U(\phi) = (1 - \phi^2)^2 \,.
\ee
Here, $V(\phi) = 1 - \phi^2$ so the superpotential is
$W(\phi) = \phi - \frac{1}{3}\phi^3$. $U$ has double (quadratic) zeros
at $\phi = \pm 1$. 

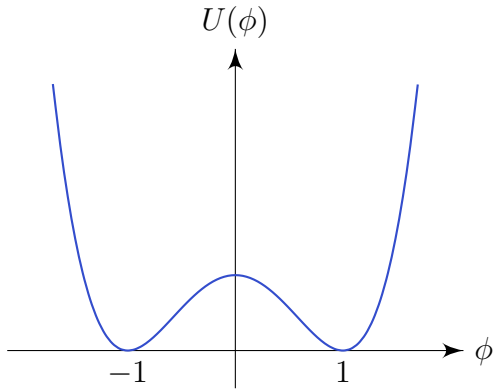
\begin{figure}
\begin{center}
  \begin{tikzpicture}
    \draw [->](-3, -1) -- (3, -1) node [right] {$\phi$};
    \draw [->] (0, -1.5) -- (0, 3) node [above] {$U(\phi)$};

    \draw [mblue, thick, domain=-2.4:2.4, samples=50] plot [smooth]
    (\x, {4 * ((\x/2)^4 - (\x/2)^2)}) ;

    \node at (-1.414, -1) [below] {$-1$};
    \node at (1.414, -1) [below] {$1$};
  \end{tikzpicture}
\end{center}
\caption{Potential of $\phi^4$-theory.}
\end{figure}

The solution of the field equation is
\be
x = \int \frac{d\phi}{1 - \phi^2} \,.
\ee
The integrand is a meromorphic differential on the $\phi$-plane, with simple
poles at $\pm 1$. The method of partial fractions gives
\bea
x &=& \half \int \left\{ \frac{1}{1 - \phi} + \frac{1}{1 + \phi}
\right\} \, d\phi \nonumber \\
&=&\half \log \left( \frac{1+\phi}{1-\phi} \right) \,,
\eea
which can be inverted to give the explicit solution (Fig.2)
\be
\phi(x) = \frac{e^{2x} - 1}{e^{2x} + 1} = \tanh x \,.
\ee
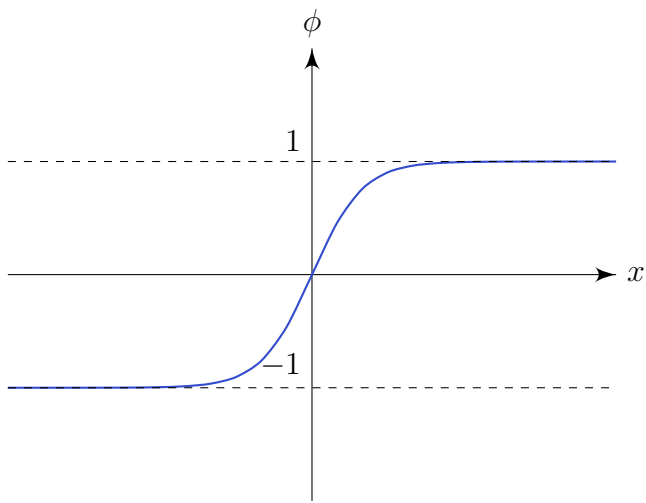
\begin{figure}
\begin{center}
  \begin{tikzpicture}
    \draw [->] (-4, 0) -- (4, 0) node [right] {$x$};
    \draw [->] (0, -3) -- (0, 3) node [above] {$\phi$};

    \draw [mblue, thick, domain=-4:4] plot [smooth] (\x, {1.5*tanh(1.5*(\x))});

    \draw [dashed] (-4, 1.5) -- (4, 1.5);
    \draw [dashed] (-4, -1.5) -- (4, -1.5);

    \node [anchor = south east] at (0, 1.5) {$1$};
    \node [anchor = south east] at (0, -1.5) {$-1$};
  \end{tikzpicture}
\end{center}
\caption{Kink of $\phi^4$-theory.}
\end{figure}
The kink interpolates between $\phi_- = -1$ and $\phi_+ = 1$, and
has short-range (exponentially decaying) tails in both directions
because the zeros of $U$ are quadratic. The kink energy is
$E = W(1)-W(-1) = \frac{4}{3}$.

The $\phi^6$-theory potential \cite{Loh}
\be
U(\phi) = \phi^2(1-\phi^2)^2
\ee
has double zeros at $-1$, $0$ and $1$. For the kink connecting $0$ to
$1$, the partial fraction method gives
\bea
x &=& \int \frac{d\phi}{\phi(1-\phi^2)} \nonumber \\ 
&=& \half \int \left\{ \frac{1}{1 - \phi} + \frac{2}{\phi}
- \frac{1}{1 + \phi} \right\} \, d\phi \nonumber \\
&=& \log \frac{\phi}{\sqrt{1 - \phi^2}} \,.
\eea
Because the coefficients of the partial fractions are commensurate,
inversion of this expression is straightforward, and the explicit
solution is
\be
\phi(x) = \frac{1}{\sqrt{1 + e^{-2x}}} \,.
\ee
Here, $V(\phi) = \phi - \phi^3$, so $W = \half\phi^2 -
\frac{1}{4}\phi^4$ and the kink energy is $\frac{1}{4}$. For the
kink connecting $-1$ to $0$, $V$ needs to be $-\phi + \phi^3$ to be positive,
and $W$ also has reversed sign. The kink energy is the same as before.

The potential \cite{Loh,GMB}
\be
U(\phi) = (a^2 - \phi^2)^2(b^2 - \phi^2)^2 \,,
\ee
with $b>a>0$, is a symmetric $\phi^8$-theory potential with four double
zeros. The field equation can be solved in a similar way, both
for the kink connecting $-a$ to $a$, and for the kink connecting
$a$ to $b$ (with the kink from $-b$ to $-a$ similar).
The kink from $-a$ to $a$ is implicitly
\be
x = \frac{1}{2a(b^2-a^2)}\log\left(\left(\frac{a+\phi}{a-\phi}\right)
  \left(\frac{b-\phi}{b+\phi}\right)^{a/b}\right) \,.
\ee
For general $a/b$ the power in this expression is not rational, and
inversion is not possible. Even if $a/b$ is rational, inversion
generally requires the solution of an algebraic equation of high
order, which is often not possible, but some cases are tractable.
$W(\phi)$ is easy to calculate for any $a$ and $b$, and the energy of
the kink is $\frac{4}{15} a^3(5b^2-a^2)$.

In the examples so far, $U$ has had only double zeros, so the kinks
have short-range tails. In the next example, $U$ is the symmetric
$\phi^8$-theory potential with one quartic zero \cite{Loh,RMGC},
\be
U(\phi) = \phi^4(1-\phi^2)^2 \,.
\ee
Using partial fractions, the kink solution
connecting $0$ to $1$ is
\bea
x &=& \int \frac{d\phi}{\phi^2(1-\phi^2)} \nonumber \\
&=& \half \int \left\{ \frac{1}{1 - \phi} + \frac{2}{\phi^2}
+ \frac{1}{1 + \phi} \right\} d\phi \nonumber \\
&=& \half\log\left(\frac{1+\phi}{1-\phi}\right) - \frac{1}{\phi} \,.
\label{phi8kink}
\eea
This illustrates the generic result that the integral of a rational
function is a combination of a rational part and a logarithmic
part \cite{Har}. These cannot usefully be combined, and there is no explicit
formula for $\phi(x)$. Instead, (\ref{phi8kink}) should be
regarded as the complete solution. The kink connects $\phi_- = 0$,
where the tail is long-range, to $\phi_+ = 1$, where the tail
is short-range. In this example, $V(\phi) = \phi^2 - \phi^4$, so
$W(\phi) = \frac{1}{3}\phi^3 - \frac{1}{5}\phi^5$ and the kink energy
is $E = W(1) - W(0) = \frac{2}{15}$. 

It is of interest to find a kink with two long-range tails,
interpolating between two quartic zeros of $U$. The simplest suitable
polynomial potential is \cite{GE}
\be
U(\phi) = (1 - \phi^2)^4 \,,
\ee
which has quartic zeros at $\phi = \pm 1$. Here, $V(\phi) =
(1-\phi^2)^2$ so $W(\phi) = \phi - \frac{2}{3}\phi^3 +
\frac{1}{5}\phi^5$. The kink connecting $-1$ to $1$ therefore has
energy $E = W(1) - W(-1) = \frac{16}{15}$. The field equation has the
solution
\be
x = \int \frac{d\phi}{(1 - \phi^2)^2} \,,
\ee
which, in terms of partial fractions, becomes
\bea
x &=& \frac{1}{4} \int \left\{ \frac{1}{(1-\phi)^2} + \frac{1}{1-\phi}
  + \frac{1}{(1+\phi)^2} + \frac{1}{1+\phi} \right\} d\phi \nonumber \\
&=& \half \frac{\phi}{1-\phi^2} + \frac{1}{4} \log \left(
    \frac{1+\phi}{1-\phi}\right) \,.
\eea
There is no explicit kink solution $\phi(x)$. However one can verify that
$\phi(x)$ approaches its asymptotic values $\pm 1$ with
long-range $\frac{1}{4x}$ tails modified by logarithmic corrections.
More general polynomial potentials with kinks of this type
have recently been comprehensively studied \cite{Bli}.

There is also a purely algebraic BPS kink with long-range tails. For this,
$U$ needs to be rational rather than polynomial. A suitable potential is
(Fig.3)
\be
U(\phi) = \frac{(1 - \phi^2)^4}{(1+\phi^2)^2} \,.
\ee
This again has quartic zeros at $\phi = \pm 1$ and is elsewhere
positive. The kink solution connecting $-1$ and $1$ is
\bea
x &=& \int \frac{1 + \phi^2}{(1 - \phi^2)^2} \, d\phi \nonumber \\
&=& \half \int \left\{ \frac{1}{(1 - \phi)^2} + \frac{1}{(1 + \phi)^2}
\right\} \, d\phi \,.
\label{AlgLongRange}
\eea
The integrand here is an Abelian differential of the second kind, with no
simple poles. The integral is therefore rational, and can be
inverted. Explicitly,
\be
x = \half\left\{\frac{1}{1-\phi} - \frac{1}{1+\phi}\right\}
= \frac{\phi}{1-\phi^2}
\ee
so the kink solution is (Fig.4)
\be
\phi(x) = \frac{\sqrt{1+4x^2} - 1}{2x} \,.
\ee
\begin{figure}
\begin{center}
  \begin{tikzpicture}
    \draw [->](-4.5, 0) -- (4.5, 0) node [right] {$\phi$};
    \draw [->] (0, -1) -- (0, 4) node [above] {$U(\phi)$};

    \draw [mblue, thick, domain=-4.0:4.0, samples=50] plot [smooth]
    (\x, {(1 - (\x/2)^2)^4) / (1 + (\x/2)^2)^2}) ;

    \node at (-2, 0) [below] {$-1$};
    \node at (2, 0) [below] {$1$};
  \end{tikzpicture}
\caption{Rational potential $U(\phi) = (1 - \phi^2)^4 / (1+\phi^2)^2$.}
\end{center}
\end{figure}
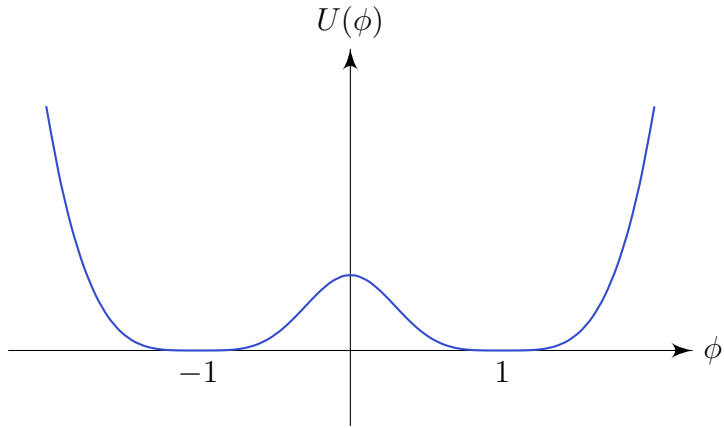

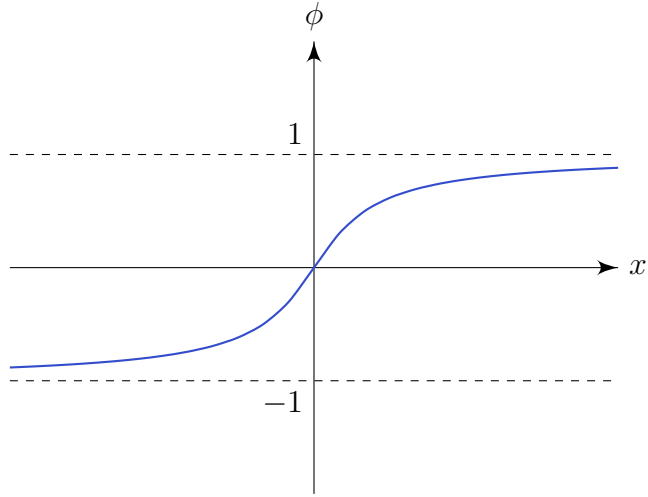
\begin{figure}
\begin{center}
  \begin{tikzpicture}
    \draw [->] (-4, 0) -- (4, 0) node [right] {$x$};
    \draw [->] (0, -3) -- (0, 3) node [above] {$\phi$};

    \draw [mblue, thick, domain=-4:4] plot [smooth]
    (\x, {(3*\x) / (1 + (sqrt(1 + 4*(\x)^2)))});

    \draw [dashed] (-4, 1.5) -- (4, 1.5);
    \draw [dashed] (-4, -1.5) -- (4, -1.5);

    \node [anchor = south east] at (0, 1.5) {$1$};
    \node [anchor = north east] at (0, -1.5) {$-1$};
  \end{tikzpicture}
\end{center}
\caption{Algebraic kink for rational potential.}
\end{figure}
$\phi(x)$ approaches $\pm 1$ asymptotically with long-range
$\frac{1}{2x}$ tails. The superpotential $W$ is more complicated
than a polynomial. As
\be
V(\phi) = \frac{(1-\phi^2)^2}{1+\phi^2} =
-3 + \phi^2 + \frac{4}{1 + \phi^2} \,,
\ee
it follows that $W(\phi) = -3\phi + \frac{1}{3}\phi^3 + 4\tan^{-1}\phi$,
and the kink energy is $E = -\frac{16}{3} + 2\pi$.

\section{Example of Semi-BPS Kink}

The only example we consider occurs for the Christ--Lee potential
\cite{CL,DGRS}
\be
U(\phi) = \half(1-\phi^2)^2 (1 + \phi^2) \,,
\ee
which has double zeros at $\pm 1$ and simple zeros at $\pm i$. $U$ has a
non-negative square root for all real $\phi$, and there is a kink
connecting $-1$ and $1$. Because $\sqrt{U}$ has just two branch
points, $\Sigma$ has genus zero, and the kink and its energy can
be determined using elementary integration techniques
(e.g. the hyperbolic substitution $\phi = \sinh y$), although the
integrals are relatively complicated. The kink solution is
\bea
x &=& \sqrt{2} \int \frac{d\phi}{(1-\phi^2)\sqrt{1+\phi^2}} \nonumber \\
&=& \frac{1}{2} \log \left(
\frac{\sqrt{2(1+\phi^2)} + 2\phi}{\sqrt{2(1+\phi^2)} - 2\phi} \right)
\,,
\eea
which can be inverted to give the explicit solution
\be
\phi(x) = \frac{ \sinh x}{ \sqrt{\sinh^2 x + 2}} \,.
\ee
By a similar integration one finds that
\bea
W(\phi) &=& \frac{1}{\sqrt{2}} \int (1-\phi^2)\sqrt{1+\phi^2}
\, d\phi \nonumber \\
&=& \frac{1}{\sqrt{2}}
\left( \frac{3}{8}\phi - \frac{1}{4}\phi^3 \right) \sqrt{1+\phi^2}
+ \frac{5}{8\sqrt{2}} \sinh^{-1}\phi \,.
\eea
The kink energy is therefore
\be
E = W(1) - W(-1) = \frac{1}{4} + \frac{5}{4\sqrt{2}} \sinh^{-1}1 \,.
\ee

\section{Examples of Sphalerons}
\vspace{4mm}

A sphaleron occurs if $U$ has a double zero (or a zero of higher even order)
$\phi_0$ and an adjacent real simple zero $\phi_1$, with $U$ positive
between them. The sphaleron solution $\phi(x)$ runs from $\phi_0$ to
$\phi_1$ and back to $\phi_0$ as $x$ runs from $-\infty$ to
$\infty$. $\phi_1$ is attained at some
finite $x$. $V=\sqrt{U}$ is not a polynomial, and is not real on the
unattained side of the simple zero, so a sphaleron is semi-BPS.
With care, the Bogomolny argument can still be used to obtain the
field equation and the sphaleron energy. If $U$ has only one more
simple zero, possibly at infinity, then the Riemann surface $\Sigma$
is of genus zero, and the field equation can be solved by
elementary integration.

The simplest example, a scaled version of what was called the
``simplest sphaleron'' in ref.\cite{MR}, has the cubic potential (Fig.5)
\cite{Ave}
\be
U(\phi) = \frac{4}{a^2}\phi^2(a^2 - \phi) \,,
\label{cubpot}
\ee
with a double zero at $\phi_0 = 0$ and a simple zero at
$\phi_1 = a^2$.
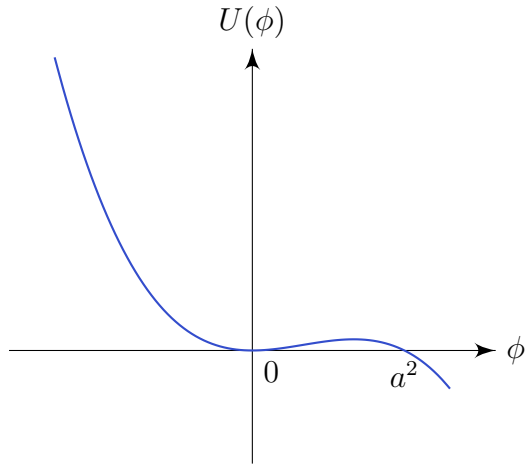
\begin{figure}
\begin{center}
  \begin{tikzpicture}
    \draw [->](-3.2, 0) -- (3.2, 0) node [right] {$\phi$};
    \draw [->] (0, -1.5) -- (0, 4) node [above] {$U(\phi)$};

    \draw [mblue, thick, domain=-2.6:2.6, samples=50] plot [smooth]
    (\x, {((\x/2)^2 - (\x/2)^3)}) ;

    \node [anchor = north west] at (0, 0) {$0$};
    \node at (2, 0) [below] {$a^2$};
  \end{tikzpicture}
\end{center}
\caption{Cubic potential with double zero and simple zero.}
\end{figure}
As $U$ is negative in the range $\phi > a^2$, and
unbounded below, the local minimum of $U$ is not the global minimum, so the
constant solution $\phi(x) = 0$ is a false vacuum with zero energy. The
sphaleron, which extends between $0$ and $a^2$ and back, represents
the energy barrier to be crossed, between the false vacuum and
configurations of negative energy. 

The algebraic curve $V^2 = U(\phi) = \frac{4}{a^2}\phi^2(a^2 - \phi)$ is a
nodal cubic curve of genus zero, and has the rational parametrisation
\be
\phi = a^2 - t^2 \,, \quad V = -\frac{2}{a} t(a^2 - t^2) \,.
\ee
$\Sigma$, the Riemann surface of the curve, is a double-cover of the
$\phi$-plane. where the single node point $V = \phi = 0$ on
the curve becomes two distinct places on $\Sigma$, with
parameter values $t = \pm a$. The sphaleron solution runs from the
node $\phi_0 = 0$ back to the node, encircling the branch point at
$\phi_1 = a^2$, so it runs along $\Sigma$ from $t = -a$ to $t= a$.

The sphaleron solution is
\be
x = \int \frac{d\phi}{V(\phi)} =
\frac{a}{2} \int \frac{d\phi}{\phi\sqrt{a^2-\phi}} \,.
\ee
Using the rational parametrisation, this becomes
\be
x = a \int \frac{dt}{a^2 - t^2}
= \tanh^{-1} \frac{t}{a} \,.
\ee
Therefore $t = a \tanh x$, and the explicit solution is (Fig.6)
\be
\phi(x) = \frac{a^2}{\cosh^2 x} \,.
\label{cubsphal}
\ee
\begin{figure}
\begin{center}
  \begin{tikzpicture}
    \draw [->] (-4, 0) -- (4, 0) node [right] {$x$};
    \draw [->] (0, -0.5) -- (0, 4) node [above] {$\phi$};

    \draw [mblue, thick, domain=-4:4] plot [smooth] (\x, {(2/cosh(\x)^2)});

    \draw [dashed] (-4, 2) -- (4, 2);

    \node [anchor = north west] at (0, 0) {$0$};
    \node [anchor = south west] at (0, 2) {$a^2$};
  \end{tikzpicture}
\end{center}
\caption{Cubic sphaleron.}
\end{figure}
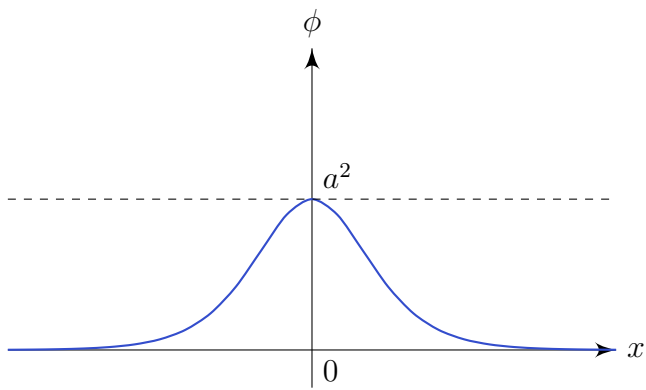
Because the potential is cubic, we call this the cubic sphaleron.
Its energy is
\be
E = \int V(\phi) \, d\phi = \frac{2}{a} \int \phi \sqrt{a^2 - \phi} \,
d\phi \,,
\ee
where the integral runs from node to node around the branch point. In
the rational parametrisation, this becomes
\be
E = \frac{4}{a} \int_{-a}^a (a^2 t^2 - t^4) \, dt = \frac{16}{15} a^4 \,.
\ee
Notice that $E$ is positive, i.e. larger than the zero energy of the
false vacuum.

The sphaleron solution (\ref{cubsphal}) is symmetric in $x$, and has a
maximum at $x=0$, so the translation zero mode of the sphaleron, its
derivative with respect to $x$, has a zero at $x=0$, i.e. has one
node. By the Sturm oscillation theorem, the second-variation operator
must have precisely one nodeless eigenfunction with negative squared
frequency -- the sphaleron's unstable mode. This argument verifies that
the solution $\phi(x)$ is a sphaleron -- an unstable static solution of the
Euler--Lagrange equation with one mode of instability.  

Although the Bogomolny argument correctly gives the field equation for
the sphaleron, and also the sphaleron energy (provided the
contours of integration are carefully chosen), this does not mean that
the sphaleron is energy-minimising. Because $\sqrt{U(\phi)}$ becomes
imaginary for $\phi > a^2$, the integrated quadratic expression in
(\ref{Bogomanip}) can be less than zero. In one direction, the unstable
deformation of the sphaleron locally makes $\phi > a^2$ and lowers the
energy. By symmetry, and rather curiously, the reversed deformation,
where $\phi$ is everywhere between $0$ and a value less than $a^2$,
also lowers the energy.

Another interesting sphaleron -- the quartic sphaleron -- occurs
for the potential
\be
U(\phi) = \frac{4}{a^2b^2} \phi^2(a^2-\phi)(b^2-\phi) \,,
\label{quartpot}
\ee
with $b>a>0$. The sphaleron runs from $\phi_0=0$ to $\phi_1 = a^2$ and
back. It represents the barrier for tunnelling from the zero-energy
false vacuum at $\phi = 0$ to the true negative-energy vacuum lying
at the potential minimum, between $a^2$ and $b^2$ \cite{CC,Ave}.

Although the potential here is quartic, it is analytically similar to the
cubic potential (\ref{cubpot}). The square root of the
quartic potential has branch points at $a^2$ and $b^2$, whereas for the
cubic potential they are at $a^2$ and infinity. To deal with
the quartic potential, we use the fractional linear transformation
\be
\Phi = \frac{(b^2-a^2)\phi}{b^2 - \phi} \,.
\ee
This maps $\phi = 0,a^2,b^2$ to $\Phi = 0,a^2,\infty$. The quartic sphaleron
solution can now be calculated in terms of $\Phi$ using the
results for the cubic sphaleron. We find, simply,
\be
\Phi(x) = \frac{a^2}{\cosh^2 x} \,,
\label{quartsphalPhi}
\ee
and in terms of the original variable $\phi$, the solution is
\be
\phi(x) = \frac{a^2b^2}{(b^2-a^2)\cosh^2 x + a^2} \,.
\label{quartsphal}
\ee

Finding the energy is algebraically more complicated than for the
cubic sphaleron, as there are extra factors from the fractional linear
transformation. However, the energy integrand is rational on the
Riemann surface $\Sigma$, and the total energy is calculated to be
\be
E = \frac{a^4}{12} \frac{1}{z^5} \left[ 
6z-4z^3+6z^5 - 3(1-z^2-z^4+z^6)\log\left(\frac{1+z}{1-z}\right)
\right] \,,
\label{quartsphalE}
\ee
where $z = \frac{a}{b}$. With the chosen normalisation factors, the
quartic potential (\ref{quartpot}) for $0 \le \phi \le a^2$ reduces
to the cubic potential (\ref{cubpot}) as $b \to \infty$, and one can
easily see that the sphaleron solution (\ref{quartsphal}) reduces to
(\ref{cubsphal}) in this limit. For the energy, the limit is
less obvious, but after expanding the logarithmic contribution in
(\ref{quartsphalE}) for small $z$, one observes several cancellations
and that the leading term in the square brackets is of order $z^5$. The
quartic sphaleron energy is $\frac{16}{15}a^4$ as $b \to \infty$, as expected.

The quartic sphaleron has an illuminating alternative form. Let $a^2 =
c^2 \tanh s$ and $b^2 = \frac{c^2}{\tanh s}$. Then (\ref{quartsphal})
becomes
\be
\phi(x) = \frac{c^2 \sinh s \cosh s}{\cosh^2 x + \sinh^2 s} \,,
\ee
which can be reexpressed as
\be
\phi(x) = \frac{c^2}{2} (\tanh(x+s) - \tanh(x-s)) \,,
\ee
that is, as a superposition of a $\phi^4$ kink centred at $-s$ and a
$\phi^4$ antikink at $s$. The kink and antikink are in unstable equilibrium
because the potential is a tilted version of the standard, symmetric
potential of $\phi^4$-theory. After a small perturbation they can
either annihilate into the false vacuum, or separate, generating an
increasing region of true vacuum.

\section{Examples of Kink Chains}
\vspace{4mm}

Kink chains are spatially periodic solutions $\phi(x)$ of the types of field
equation we have been considering. More precisely, they are
chains of alternating kinks and antikinks, with $\phi$
spatially oscillating between neighbouring simple zeros of $U(\phi)$. 
Because $U$ has simple zeros, $\Sigma$ is connected, and the kink chain
is semi-BPS. We shall not discuss kink chain solutions in much
detail. Most of them involve elliptic or hyperelliptic integrals.

The simplest kink chain arises from the quadratic potential (Fig.7)
\be
U(\phi) = 1 - \phi^2 \,.
\label{quadpot}
\ee
This is interesting, even though the Euler--Lagrange equation (\ref{EulerLag})
is the elementary linear equation $\frac{d^2\phi}{dx^2} + \phi = 0$.
\begin{figure}
\begin{center}
  \begin{tikzpicture}
    \draw [->](-3.2, 0) -- (3.2, 0) node [right] {$\phi$};
    \draw [->] (0, -1.5) -- (0, 4) node [above] {$U(\phi)$};

    \draw [mblue, thick, domain=-2.4:2.4, samples=50] plot [smooth]
    (\x, {3*(1 - (\x/2)^2)}) ;

    \node [anchor = north west] at (0, 0) {$0$};
    \node [anchor = north west] at (2, 0) {$1$};
    \node [anchor = north east] at (-2, 0) {$-1$};
  \end{tikzpicture}
\end{center}
\caption{Quadratic potential with simple zeros.}
\end{figure}
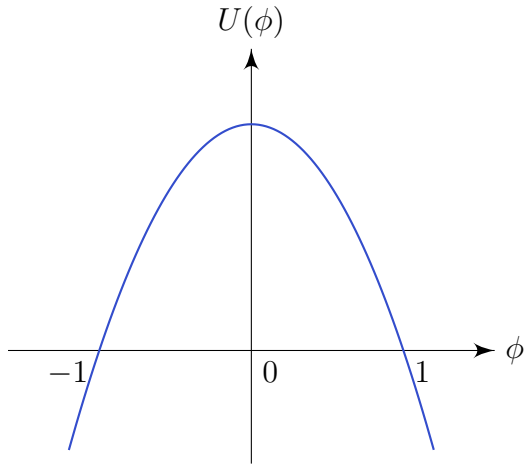
The kink chain solution is
\be
x = \int \frac{d\phi}{\sqrt{1 - \phi^2}} = \sin^{-1} \phi \,,
\ee
so
\be
\phi(x) = \sin x \,,
\ee
with spatial period $2\pi$ (Fig.8). Kinks are centred at $x=0$
${\rm mod} \ 2\pi $ and antikinks at $x = \pi$ ${\rm mod} \ 2\pi$.
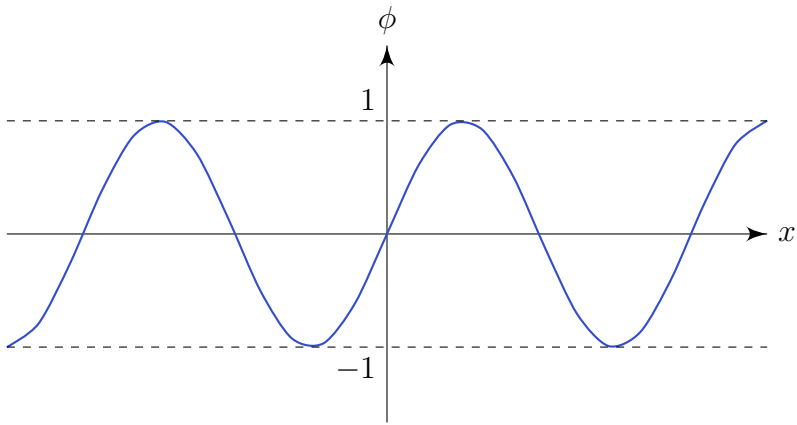
\begin{figure}
\begin{center}
  \begin{tikzpicture}
    \draw [->] (-5, 0) -- (5, 0) node [right] {$x$};
    \draw [->] (0, -2.5) -- (0, 2.5) node [above] {$\phi$};

    \draw [mblue, thick, domain=-5:5] plot [smooth] (\x, {1.5*sin(90*\x)});

    \draw [dashed] (-5, 1.5) -- (5, 1.5);
    \draw [dashed] (-5, -1.5) -- (5, -1.5);

    \node [anchor = south east] at (0, 1.5) {$1$};
    \node [anchor = north east] at (0, -1.5) {$-1$};
  \end{tikzpicture}
\end{center}
\caption{Kink chain for quadratic potential.}
\end{figure}
The energy per period is
\be
E = \oint \sqrt{1 - \phi^2} \, d\phi = \pi \,,
\ee
where the integral is along a closed loop in $\Sigma$, enclosing both
branch points. Like the cubic sphaleron, this kink chain is rather unphysical
because the potential is unbounded below. However, it often occurs for
limiting parameter values of more realistic potentials. 

A more physical kink chain occurs for the potential \cite{MS}
\be
U(\phi) = (1 - \phi^2)(1 - k^2\phi^2) \,,
\ee
with $0 < k < 1$. The solution of the field equation is the elliptic integral
\be
x = \int \frac{d\phi}{\sqrt{(1-\phi^2)(1-k^2\phi^2)}} \,,
\ee
with the integral circulating around the branch points at $\phi = \pm
1$ as $x$ increases, If we fix $\phi = 0$ and $\frac{d\phi}{dx} > 0$ at $x=0$,
then
\be
x = \int_0^\phi \frac{d\phi}{\sqrt{(1-\phi^2)(1-k^2\phi^2)}} \,.
\ee
This Legendre elliptic integral of the first kind can be inverted
using the Jacobi elliptic function ${\rm sn}$ with modulus $k$, giving
the kink chain
\be
\phi(x) = {\rm sn}(x,k) \,.
\label{Jacobichain}
\ee
$\phi$ oscillates between $-1$ and $1$, as for the quadratic potential
(\ref{quadpot}), and has equally-spaced kinks and antikinks. The
full spatial period is
\be
X = 4 \int_0^1 \frac{d\phi}{\sqrt{(1-\phi^2)(1-k^2\phi^2)}} = 4K(k) \,,
\ee
where
\be
K(k) = \frac{\pi}{2} \left(1 + \frac{1}{4} k^2 + \frac{9}{64} k^4
+ \cdots \right)
\ee
is the complete elliptic integral of the first kind.
The energy per period of the kink chain is 
\bea
E &=& 4\int_0^1 \sqrt{(1-\phi^2)(1-k^2\phi^2)} \, d\phi \nonumber \\
&=& \frac{4}{3k^2} \left((1 + k^2)E(k) - (1 - k^2)K(k) \right) \,,
\eea
where
\be
E(k) =  \frac{\pi}{2} \left(1 - \frac{1}{4} k^2 - \frac{3}{64} k^4
  - \cdots \right)
\ee
is the complete elliptic integral of the second kind.
As $k \to 0$, the potential $U$ approaches the simple quadratic form
(\ref{quadpot}), and unsurprisingly, the kink chain solution
${\rm sn}(x,k)$ approaches $\sin x$, its period $X$
approaches $2\pi$ and its energy per period approaches $\pi$.

Kink chains are not stable. The translation zero mode (the spatial
derivative of the chain) oscillates with the same period as the chain
itself, and has two nodes per period. An infinitely-long
chain therefore has infinitely many unstable modes. Ref. \cite{MS},
discussed the modes of instability for a kink chain (\ref{Jacobichain})
of finite length, having $N$ complete periods and periodic boundary
conditions. There are $2N-1$ modes of instability, all arising from
the breakdown of the equal spacing of the $N$ kinks and $N$ antikinks.
The kinks and antikinks tend to approach each other and annihilate.

\section{Trigonometric Potentials}
\vspace{4mm}

In a number of scalar field theories, in particular, the well-known
sine-Gordon theory, the potential $U(\phi)$ is a periodic,
trigonometric function of $\phi$. We will consider here the class of
potentials $U(\phi) = \cos^n \phi$, with $n$ a positive integer, and
will focus on the simplest cases $n=1,2,4$ and $8$.
For all $n$, $U = 1$ at $\phi = 0$, and the closest zeros of $U$
are at $\phi = \pm \frac{\pi}{2}$. There are static solutions
interpolating between these latter values of $\phi$. The $n=1$
potential $\cos\phi$ has simple zeros at $\pm \frac{\pi}{2}$, so the
solution is a semi-BPS kink chain. For $n=2k$, $U(\phi)$ has the real
square root $V(\phi)=\cos^k\phi$ which can be straightforwardly integrated
to obtain a superpotential $W(\phi)$, and here the solution
is a BPS kink. The minima of $U$ at $\pm \frac{\pi}{2}$ are quadratic
for $n=2$, so the kink has tails that are short-range. This is, in
fact, a variant of the sine-Gordon kink \cite{PS,Rub}. For larger,
even $n$, the minima at $\pm \frac{\pi}{2}$ are of higher order, so
the kink has long-range tails \cite{MGGL}.

For all $n$, it is possible to use the rational parameter $t = \tan
\frac{\phi}{2}$, with $t$ between $-1$ and $1$. This sometimes converts the 
integrals specifying the solutions and their energies into integrals 
that have appeared earlier. Generally, for 
$U(\phi) = \cos^n\phi$, the kink or kink chain solution is given by
\be
x = \int \frac{d\phi}{\cos^{\frac{n}{2}}\phi} \,.
\label{trigsoln}
\ee
Using $t = \tan \frac{\phi}{2}$ this becomes
\be
\frac{x}{2} = \int \frac{(1+t^2)^{\frac{n}{2} - 1}}{(1 - t^2)^{\frac{n}{2}}}
\, dt \,,
\label{ratsoln}
\ee
which has a rational integrand when $n$ is even.

For $n=1$ the solution (\ref{ratsoln}) is the
lemniscate elliptic integral of the first kind
\be
\frac{x}{2} = \int \frac{dt}{\sqrt{1-t^4}} \,,
\ee
so $t = {\rm sl}\left(\frac{x}{2}\right)$, where ${\rm sl}$ is the
lemniscate elliptic function, related to Jacobi functions via
\be
{\rm sl}(z) = {\rm sn}(z,i) = {\rm sc}(z,\sqrt{2}) \,. 
\ee
The explicit semi-BPS kink chain solution is therefore \cite{PS}
\be
\tan \frac{\phi}{2} = {\rm sl}\left(\frac{x}{2}\right) \,,
\ee
which oscillates between $\phi = \pm \frac{\pi}{2}$.

The spatial period and energy per period of this kink chain are best
found using the Beta function, whose trigonometric integral expression
and value are
\be
B(z_1,z_2) = 2 \int_0^{\frac{\pi}{2}} (\sin \phi)^{2z_1 - 1} (\cos
\phi)^{2z_2 - 1} \, d\phi = \frac{\Gamma(z_1)\Gamma(z_2)}{\Gamma(z_1+z_2)} \,.
\ee
The period $X$ and energy per period $E$ are then
\bea
X &=& 4\int_0^{\frac{\pi}{2}} \frac{d\phi}{\sqrt{\cos\phi}} =
2 \frac{\Gamma(\half)\Gamma(\frac{1}{4})}{\Gamma(\frac{3}{4})} \,, \nonumber \\
E &=& 4\int_0^{\frac{\pi}{2}} \sqrt{\cos\phi} \, d\phi
= 2 \frac{\Gamma(\half)\Gamma(\frac{3}{4})}{\Gamma(\frac{5}{4})} \,.
\eea
Their approximate values are $X \simeq 10.488$, $E \simeq 4.793$, and
there is the curious exact relation $XE = 16\pi$. 

For $n=2$, the potential is $U(\phi) = \cos^2\phi
= \half(\cos(2\phi) + 1)$, a variant of the sine-Gordon potential \cite{PS}.
The sine-Gordon kink solution in this case is
\be
x = \int \frac{d\phi}{\cos\phi} = 2 \int \frac{dt}{1-t^2} = 2
\tanh^{-1} t \,,
\ee
and therefore
\be
t = \tan{\frac{\phi}{2}} = \tanh{\frac{x}{2}} \,.
\ee
More simply,
\be
\tan\phi = \sinh x \,.
\ee
The kink connects $-\frac{\pi}{2}$ to $\frac{\pi}{2}$ as $x$ runs
from $-\infty$ to $\infty$, and the tails are short-range. Here,
$V(\phi)=\cos\phi$, so the superpotential is $W(\phi)=\sin\phi$.
The sine-Gordon kink is therefore BPS, and its energy is
\be
E = W\left(\frac{\pi}{2}\right) - W\left(-\frac{\pi}{2}\right) = 2 \,.
\ee 

For the $n=4$ potential $U(\phi)=\cos^4\phi$, the BPS kink solution
and its energy are
\be
x = \int \frac{d\phi}{\cos^2\phi} \,, \quad
E = \int_{-\frac{\pi}{2}}^{\frac{\pi}{2}} \cos^2\phi \, d\phi \,.
\ee
Both integrals are straightforward, and give results \cite{Moh}
\be
x = \tan\phi \,, \quad E = \half \pi \,,
\ee
so $\phi(x) = \tan^{-1} x$. The tail behaviours are of the long-range
form $\phi \sim \pm \frac{\pi}{2} - \frac{1}{x}$ as $x \to \pm\infty$.  
In this example, the rational parametrisation gives
\be
x = 2\int \frac{1+t^2}{(1-t^2)^2} \, dt = \frac{2t}{1-t^2} = \tan\phi \,.
\ee
The integral here is the same as in (\ref{AlgLongRange}), defining
the algebraic BPS kink solution that connects two quartic zeros
without log terms.

Similarly, for the $n=8$ potential $U(\phi) = \cos^8\phi$, the kink
solution and its energy are
\be
x = \int \frac{d\phi}{\cos^4\phi} \,, \quad
E = \int_{-\frac{\pi}{2}}^{\frac{\pi}{2}} \cos^4\phi \, d\phi \,,
\ee
so
\be
x = \tan\phi + \frac{1}{3} \tan^3\phi \,, \quad
E = \frac{3}{8} \pi \,,
\ee
where we used the identity $\sec^4\phi = (1 + \tan^2\phi)\sec^2\phi$.
The kink solution here is implicit, unless one solves a cubic
equation. The tails are dominated by the $\tan^3\phi$ term,
and approach the limiting values more slowly than in the $n=4$ case. 
With further effort, solutions for any $n$ can be determined.

\section{Conclusions}
\vspace{4mm}

We have reviewed the topologically interesting, static solutions of scalar
field theories in one dimension having a single scalar field $\phi$.
These solutions are kinks, sphalerons and kink chains, whose nature
depends critically on the multiplicities of the zeros of the potential
$U(\phi)$ appearing in the first-order field equation (the first
integral of the static Euler--Lagrange equation, with the constant of
integration absorbed into $U$).

Although the Bogomolny rearrangement
of the field energy (the completion of the square) is formally valid
for all these types of solution, we have made a novel distinction
between BPS and semi-BPS solutions. If $\sqrt{U(\phi)}$ is defined on
the $\phi$-plane, i.e. if $\sqrt{U}$ is a simple function of $\phi$
with no explicit square roots remaining, then a kink solution can be
regarded as a BPS kink. If $\sqrt{U}$ has branch points and
is well-defined only on the double-cover of the $\phi$-plane, then we
refer to a kink solution as semi-BPS. The field equation is generally
more difficult to solve in the semi-BPS case, and the energy is more tricky to
calculate. Sphaleron and kink chain solutions require $U$ to have
simple zeros (or possibly, zeros of higher odd orders); $\sqrt{U}$ therefore
has branch points and these solutions are inevitably semi-BPS. BPS and
semi-BPS kinks are stable, or at least metastable, but sphaleron and
kink chain solutions have modes of instability. We have presented a
broad range of examples of these solution types. Most are well-known,
but not all.

\vspace{4mm}

\section*{Acknowledgements}

This work has been partially supported by consolidated grant
ST/T000694/1 from the UK Science and Technology Facilities Council.
I'm grateful to the organisers of the 2023 SIG XI workshop in Krakow,
Poland, and in particular to Andrzej Wereszczynski for discussions about
the distinction between BPS and semi-BPS solutions. Ref.\cite{Izq} was
in preparation at the same time as this paper, and I am grateful for
receiving an early draft.

\end{document}